\documentclass[a4paper,12pt]{article}
\usepackage{amssymb}
\usepackage{amsfonts}
\usepackage{amsmath}
\usepackage{epsfig}
\newcommand{\be}{\begin{equation}}
\newcommand{\ee}{\end{equation}}
\newcommand{\bd}{\begin{displaymath}}
\newcommand{\ed}{\end{displaymath}}
\newcommand{\bea}{\begin{eqnarray}}
\newcommand{\eea}{\end{eqnarray}}

\begin{document}

\baselineskip 24pt
\newcommand{\sheptitle}
{\Large Perturbative Estimates of Lepton Mixing Angles \\
in Unified Models}

\newcommand{\shepauthor}
{Stefan Antusch$^1$, Stephen F. King$^2$ and Michal
Malinsk\'{y}$^3$}

\newcommand{\shepaddress}
{$^1$Max-Planck-Institut f\"ur Physik (Werner-Heisenberg-Institut),\\
F\"ohringer Ring 6, 80805 M\"unchen, Germany\\[0.2cm]
$^2$School of Physics and Astronomy, University of Southampton,
Southampton, SO17 1BJ, UK \\[0.2cm]
$^3$Department of Theoretical Physics, School of Engineering Sciences, \\
Royal Institute of Technology (KTH) -- AlbaNova University Center, \\
Roslagstullsbacken 21, SE-106 91 Stockholm, Sweden
}

\newcommand{\shepabstract}
{Many unified models predict two large neutrino mixing angles,
with the charged lepton mixing angles being small and quark-like,
and the neutrino masses being hierarchical. Assuming this, we
present simple approximate analytic formulae giving the lepton
mixing angles in terms of the underlying high energy neutrino
mixing angles together with small perturbations due to both
charged lepton corrections and renormalisation group (RG) effects,
including also the effects of third family canonical normalization
(CN). We apply the perturbative formulae to the ubiquitous case of
tri-bimaximal neutrino mixing at the unification scale, in order
to predict the theoretical corrections to mixing angle predictions
and sum rule relations, and give a general discussion of all
limiting cases.}

\begin{titlepage}
\begin{flushright}
\end{flushright}

\begin{center}
{\large{\bf \sheptitle}}
\bigskip \\ \shepauthor \\ \mbox{} \\ {\it \shepaddress} \\ \vspace{.5in}
{\bf Abstract} \bigskip \end{center} \setcounter{page}{0}
\shepabstract
\begin{flushleft}
\end{flushleft}
\end{titlepage}

\section{Introduction}
It is one of the goals of theories of particle physics beyond the
Standard Model to predict quark and lepton masses and mixings.
Whilst the quark mixing angles are known to all be rather small
\cite{Amsler:2008zz}, by contrast two of the lepton mixing angles
are identified as being rather large \cite{Schwetz:2008er}. This
observation, together with the smallness of neutrino masses,
provides a tantalizing clue in the search for the origin of quark
and lepton flavour. One possibility, widely studied in the
literature, is that these two observations are related due to the
underlying nature of neutrinos, which, unlike charged lepton and
quark masses, are electrically neutral and so may have Majorana
masses. The origin of the neutrino Majorana masses, being
different from that of the quarks and charged leptons, may then be
responsible for both the smallness of neutrino masses and the
largeness of two of the lepton mixing angles. Whilst not a
theorem, the plausibility and attractiveness of this hypothesis
makes the conclusion that the origin of the large lepton mixing
lies in the neutrino sector hard to resist. This idea is
reinforced in the framework of many (but not all) grand unified
theories (GUTs) \cite{Georgi:1974sy} where the quarks and leptons
are treated on the same footing, resulting typically in small
quark and charged lepton mixing angles 
(possibly related to each other) with the see-saw mechanism 
\cite{Minkowski:1977sc,s2,s3,s4,s5} responsible for both
small neutrino masses and large neutrino mixing angles
(see e.g.\ \cite{Antusch:2004gf,King:1998jw,King:1999mb,King:2006hn}).

Motivated by such considerations, here we shall assume
that the large lepton mixing angles originate from the neutrino
sector, and that the charged lepton mixing angles are rather
small, and have a similar pattern to the quark mixing angles.
Indeed, in many (but not all) GUTs, the origin of the quark mixing angles
derives predominantly from the down quark sector, which in turn is
closely related to the charged lepton sector. In order to
reconcile the down quark and charged lepton masses, simple
ansatze, such as the Georgi-Jarlskog hypothesis \cite{Georgi:1979df}, lead to very
simple approximate expectations for the charged lepton mixing
angles such as $\theta^e_{12}\approx \lambda/3$,
$\theta^e_{23}\approx \lambda^2$, $\theta^e_{13}\approx
\lambda^3$, where $\lambda \approx 0.22$ is the Wolfenstein
parameter \cite{Amsler:2008zz} from the quark mixing matrix. Although the charged
lepton mixing angles are clearly expected to be rather small,
nevertheless it is important to take into account such charged
lepton corrections in order to estimate reliably the physical
lepton mixing angles (see for example\ \cite{Antusch:2005kw}).

Another effect which must be taken into
account is the renormalisation group (RG) running required to
relate high energy (GUT scale) predictions to low energy neutrino
experiments.
It is typically calculated numerically by solving the relevant
coupled system of renormalisation group equations including
the one for the effective neutrino mass matrix  \cite{Chankowski:1993tx,Babu:1993qv,Antusch:2001ck,Antusch:2001vn,Antusch:2002ek}
and taking into account the mass thresholds of the right-handed neutrinos \cite{King:2000hk,Antusch:2002rr,Antusch:2005gp}.
In many GUT models the neutrino masses turn out to be
hierarchical in nature, and in such cases the RG running effects
are relatively small (see e.g.\ \cite{Haba:1999fk}), but none the less such effects may be
competitive with the charged lepton corrections and so also must
be taken into account before comparing GUT scale predictions to
low energy experiment.
In this case analytic approximations for the RG corrections to the
neutrino parameters can be used (see e.g.\ \cite{Antusch:2005gp,Chankowski:1999xc,Casas:1999tg,Antusch:2003kp}).
A third class of correction, not so well
studied or appreciated, but nevertheless important in realistic
models, are the canonical normalization (CN) effects resulting
from the kinetic terms receiving corrections from the same physics
responsible for the generation of flavour. Although model
dependent, we have shown \cite{Antusch:2007ib,Antusch:2007vw} that the dominant
canonical normalization correction
arising due to the physics responsible for the third family Yukawa
couplings (more precisely from dominant 33-elements of the charged
lepton and neutrino Yukawa matrices in the flavour basis)
has the same structure as the leading logarithmic (log) RG
corrections, and so both effects may be subsumed into a single
parameter $\eta$. To be precise, $\eta= \eta^{RG}+\eta^{CN}$, where in the MSSM
\be
\eta^{RG} \approx \frac{y_{\tau}^2}{8\pi^2}\ln \frac{M_{GUT}}{M_Z}
+\frac{y_{\nu_3}^2}{8\pi^2}\ln \frac{M_{GUT}}{M_3}
\label{etaRG}
\ee
and $y_{\tau}$ is the tau Yukawa coupling, while $y_{\nu_3}$ is the
largest neutrino Yukawa coupling associated with a heavy
right-handed neutrino mass threshold $M_3$, with $M_{GUT}$
being the GUT scale (see also section 6.1 of \cite{Antusch:2005gp}). The parameter $\eta^{CN}$ parametrises
CN effects and is highly model dependent, however it
contributes in the same way as $\eta^{RG}$ to leading log, since (in
supersymmetric theories) both
effects arise from third family wavefunction corrections
in the considered approximation. Therefore the combined
effects of RG corrections and CN effects
will be approximately parametrized by a single parameter
$\eta$ in the analytic estimates which follow.

In this paper we provide simple approximate analytic formulae
giving the lepton mixing angles and phases
in terms of the neutrino mixing angles and phases
together with perturbative corrections due to non-zero charged lepton mixing angles and phases,
leading log renormalisation group running corrections
and third family canonical normalization effects.
We derive such approximate analytic perturbative corrections
to leading order in the charged lepton
mixing angles $\theta_{ij}^e$ and the RG/CN universal parameter $\eta$,
where these parameters are all assumed to be small as
discussed above. The resulting expansions
provide useful physical insight into the origin and
nature of the deviations of the observable lepton mixing angles from
the underlying neutrino mixing angles at the GUT scale.
In addition such perturbative formulae
may be useful for speeding up multi-parameter scans in particular
GUT models, or simply as a means of checking the numerical
results. With the additional assumption that the underlying
high energy {\it neutrino} (but not the physical lepton) mixing angles have the
tri-bimaximal (TB) form \cite{Harrison:2002er}, we use the perturbative formulae to
derive new relations between lepton mixing angles
and the perturbative charged lepton and RG/CN corrections.

The layout of the remainder of the paper is as follows.
In Section \ref{conv} we state the main conventions used in the paper.
In Section \ref{sec:formulae} we present the analytic formulae for the
lepton mixing angles in terms of the underlying high energy neutrino
mixing angles together with small perturbations due to both
charged lepton corrections and RG/CN effects.
In Section \ref{deviation} we give the parameterization of the
lepton mixing angles in terms of parameters describing the
deviations from tri-bimaximal mixing. In Section \ref{TBMsum}
we specialize to the case of tri-bimaximal {\em neutrino} mixing,
and give the perturbative formulae in this case. We then go on to apply these
results first to the case of GUT models, and then discuss the results for
various limiting cases. Section \ref{concl} concludes the paper.

\section{Conventions}\label{conv}

The mixing matrix in the lepton sector, the PMNS matrix
$U_{PMNS}$, is defined as the matrix appearing in the electroweak
coupling to the $W$ bosons expressed in terms of lepton mass
eigenstates. The Lagrangian is given in terms of mass matrices of
charged leptons $M_e$ and neutrinos $m_{LL}$ as
\begin{equation}
\mathcal{L}=-\overline{e_L} M_e e_R -\frac{1}{2} \overline{\nu_L}
m_{LL}\nu_L^c + H.c. \;.
\end{equation}
The change from the flavour basis to the mass eigenbasis is achieved via
\begin{equation}
V_{e_L} M_e V_{e_R}^\dagger = diag(m_e, m_\mu, m_\tau),
\hspace{3mm} V_{\nu_L} m_{LL} V_{\nu_L}^T = \mbox{diag}(m^{\nu}_1, m^{\nu}_2, m^{\nu}_3),
\end{equation}
where $V_{e_{L}}$, $V_{e_{R}}$ and $V_{\nu_{L}}$ are $3\times 3$ unitary matrices.
The PMNS matrix (in the ``raw'' form, i.e.\ before the ``unphysical'' phases were absorbed into redefinitions of the relevant lepton field operators) is then given by
\begin{equation}
U_{PMNS} =V_{e_L} V_{\nu_L}^\dagger. \label{PMNS}
\end{equation}

We use the parameterization $ U_{\mathrm{PMNS}} = U_{23} U_{13}
U_{12} $ with $U_{23}, U_{13}, U_{12}$ being defined as
\bea
U_{12}= \left(\begin{array}{ccc}
  c_{12} & s_{12}e^{-i\delta_{12}} & 0\\
  -s_{12}e^{i\delta_{12}}&c_{12} & 0\\
  0&0&1\end{array}\right)
  , \:&
\quad U_{13}=\left(\begin{array}{ccc}
   c_{13} & 0 & s_{13}e^{-i\delta_{13}}\\
  0&1& 0\\
  - s_{13}e^{i\delta_{13}}&0&c_{13}\end{array}\right)  ,  \nonumber
  \eea
  \bea
  \label{eq:U23U13U12}
U_{23}=\left(\begin{array}{ccc}
 1 & 0 & 0\\
0&c_{23} & s_{23}e^{-i\delta_{23}}\\
0&-s_{23}e^{i\delta_{23}}&c_{23}
 \end{array}\right)
\eea
where $s_{ij}$ and $c_{ij}$ stand for $\sin \theta_{ij}$ and
$\cos \theta_{ij}$, respectively and the remaining 3 unphysical phases have been rotated away,
see for instance\ \cite{King:2002nf} for further details. The same scheme shall be used for the individual charged lepton and neutrino sector rotations in (\ref{PMNS}), with superscripts at the relevant quantities.
Recall that in the standard PDG parameterisation \cite{Amsler:2008zz} the Dirac CP phase $\delta$ relevant for neutrino oscillations and the Majorana CP phases $\alpha_1$ and $\alpha_2$ are entering as follows:
\be
U_{PMNS}=R_{23}U_{13}R_{12}P_{0}\label{UPDG}\;,
\ee
where $P_{0}$ is a complex diagonal matrix $P_{0}={\rm diag}(e^{i\alpha_1/2},e^{i \alpha_2/2},1)$ and
\bea
R_{23}=\left(\begin{array}{ccc}
 1 & 0 & 0\\
0&c_{23} & s_{23}\\
0&-s_{23}&c_{23}
 \end{array}\right),\;\;
 U_{13}=\left(\begin{array}{ccc}
   c_{13} & 0 & s_{13}e^{-i\delta}\\
  0&1& 0\\
  - s_{13}e^{i\delta}&0&c_{13}\end{array}\right)\;,\;\;
  R_{12}= \left(\begin{array}{ccc}
  c_{12} & s_{12} & 0\\
  -s_{12}&c_{12} & 0\\
  0&0&1\end{array}\right).
\eea
Comparing (\ref{PMNS}) and (\ref{eq:U23U13U12}) with (\ref{UPDG}) one finds $\delta = \delta_{13}-\delta_{23}-\delta_{12}$,
$\alpha_2 = 2 \delta_{23}$ and $\alpha_1 =2(\delta_{12}+\delta_{23})$ \cite{King:2002nf}
(after having performed irrelevant global rephasings to absorb the ``unphysical'' phases).
The mixing angles $\theta_{ij}$ are the same in both parameterizations.

\section{Charged lepton and RG/CN perturbations}\label{sec:formulae}
In the considered GUT motivated framework defined above, with
hierarchical neutrino masses
$\frac{m_{2}^{\nu}}{m_{3}^{\nu}}\approx \sqrt{\frac{\Delta
m^{2}_{\odot}}{|\Delta m^{2}_{A}|}}\approx 0.18$, the low energy
lepton mixing angles are dominated by the high energy neutrino
sector contributions which are subject to three classes of
perturbations: 1) canonical normalization effects due to a
would-be non-canonical structure of the kinetic terms emerging at
the GUT-scale,\; 2) contributions from the charged lepton mixings
$\theta_{ij}^e$ and 3) the RG corrections. The goal of this section
is to derive formulae for the lepton mixing angles and phases in terms of the
neutrino mixing angles and phases, together with small perturbative
corrections due to the above three effects.

Let us first discuss the canonical normalization effects. In order
to get the correct asymptotic behaviour of the matter sector
propagators one should first bring the relevant kinetic terms into
the canonical form by a suitable field redefinition $\hat L_{L}\to
P_{L}^{-1}\hat L_{L}\equiv L_{L}$, $\hat e_{R}\to
P_{e_{R}}^{-1}\hat e_{R}\equiv e_{R}$ where $L_{L}$ stands for the
$SU(2)_{L}$ lepton doublet $(\nu_{L},e_{L})$ and hats denote for
the corresponding quantities in the defining basis (i.e.\.  before
canonical normalization). This, however, affects also the defining
basis mass matrices $\hat M_{e}$ and $\hat m_{LL}$ as follows:
$\hat M_{e}\to P_{L}^{T}\hat M_{e}P_{e^{c}}\equiv M_{e}$ and $\hat
m_{LL}\to P_{L}^{T}\hat m_{LL}P_{L}\equiv m_{LL}$\;. The charged lepton mass matrix $M_e$ and the effective neutrino mass 
matrix $m_{LL}$ \footnote{Above the seesaw scales, $m_{LL}$ refers
to the combination of parameters $v^2 Y_\nu M_{RR}^{-1} Y_\nu^T$, 
where $Y_\nu$ is the running neutrino Yukawa matrix, $M_{RR}$ 
the running mass matrix of the right-handed neutrinos and $v$ is the 
low scale value of the VEV of the Higgs which is involved in the neutrino 
Yukawa interactions.} 
are subsequently evolved to the low scale my means of the renormalisation 
group equations and the unitary transformations $V_{e_{L}}$ and
$V_{\nu_{L}}$ entering formula (\ref{PMNS}) can be extracted.

Turning to the RG effects, as we pointed out in \cite{Antusch:2007ib,Antusch:2007vw},
if third family contributions dominate both CN and RG corrections, the
CN and RG effects can be subsumed (at leading log) into a single parameter
$\eta$ denoting the non-universality in the 33 component of the
$P_{L}$ matrix. An interested reader can find the technical
details of how to obtain the low-scale diagonalisation matrices
$V_{e_{L}}$ and $V_{\nu_{L}}$ given their GUT-scale counterparts
$\hat V_{e_{L}}$ and $\hat V_{\nu_{L}}$ elsewhere
\cite{Antusch:2007ib,Antusch:2007vw}
(in particular see Eqs.(2.14) and (2.17) of \cite{Antusch:2007vw})
although we emphasize that the resulting lepton mixing angles
and phases were not explicitly expanded in terms of $\eta $
as we do here.

The formulae for the lepton mixing angles and phases $\theta_{ij}$, $\delta_{ij}$
in terms of the neutrino mixing angles and phases $\theta_{ij}^{\nu}$, $\delta_{ij}^{\nu}$,
together with small perturbative corrections due
to charged lepton mixing angles and phases $\theta_{ij}^{e}$, $\delta_{ij}^{e}$
have already appeared in the literature
\cite{Antusch:2005kw,King:2005bj}. The new physics that we wish to
discuss here is the effect of the additional perturbative CN/RG corrections described by the
universal parameter $\eta$. With $\eta$ included, using the techniques described above,
the leading order expansions for the physical lepton
mixing angles and phases in terms of the relevant neutrino and charged lepton sector quantities
$\theta_{ij}^{\nu,e}$, $\delta_{ij}^{\nu,e}$ become:
\bea \label{F1}
{s_{23}}e^{-i\delta_{23}}&\approx&
{s_{23}^{\nu}}\left(1+\frac{\eta}{2}{c_{23}^{\nu}}^{2}\right)e^{-i\delta_{23}^{\nu}}
-{\theta_{23}^{e}}{c_{23}^{\nu}}e^{-i\delta_{23}^{e}}\;,
\\
\label{F2} {s_{13}}e^{-i\delta_{13}}&\approx&
{\theta}_{13}^{\nu}e^{-i\delta_{13}^{\nu}}
-{\theta_{12}^{e}}{s_{23}^{\nu}}e^{-i(\delta_{23}^{\nu}+\delta_{12}^{e})}+
 \frac{m_{2}^{\nu}}{m_{3}^{\nu}}\eta c_{12}^{\nu}s_{12}^{\nu}c_{23}^{\nu}s_{23}^{\nu}
\, e^{-i(\delta_{12}^{\nu}-\delta_{23}^{\nu})}\;,
\\
\label{F3} {s_{12}}e^{-i\delta_{12}}&\approx&
{s_{12}^{\nu}}\left(1+\frac{\eta}{2}{c_{12}^{\nu}}^{2}{s_{23}^{\nu}}^{2}\right)e^{-i\delta_{12}^{\nu}}
-{\theta_{12}^{e}}{c_{23}^{\nu}}{c_{12}^{\nu}}e^{-i\delta_{12}^{e}}\;.
\eea
They should be compared to the results with only charged lepton corrections
included \cite{Antusch:2005kw,King:2005bj}, to which these results reduce in the limit $\eta =0$.
We have neglected $\theta_{13}^e$ since in GUT models
we expect that $\theta_{13}^e \approx \lambda^3$ and so $\theta_{13}^e$ terms
may be regarded as higher order. We have included terms like
$\frac{m_{2}^{\nu}}{m_{3}^{\nu}}\eta$ which may compete with $\theta_{12}^e\approx \lambda /3$,
and have also included terms like $\theta_{23}^e\approx \lambda^2$ which are not so different
from $\theta_{12}^e\approx \lambda /3$.
Terms of the order ${\cal O}(m_1/m_2)$ and ${\cal O}(m_1/m_3)$ have been neglected,
which corresponds to the assumption of a strong  hierarchy of the neutrino mass spectrum. This also implies $\frac{m^\nu_2}{m^\nu_3} \approx 
\sqrt{\frac{\Delta m_{\odot}^{2}}{|\Delta m_{A}^{2}|}}$\, which is a quantity directly accessible in the neutrino oscillation experiments.

We would like to remark that, in general, one of the main sources of errors
in the leading log approximation for the RG corrections is associated to the fact that
the 3rd family Yukawa couplings (i.e.\ $y_\tau$ and $y_{\nu_3}$)
are themselves running quantities. However, since the running of
$y_\tau$ and $y_{\nu_3}$ affects the corrections to all the mixing
parameters in the same way, effectively only modifying
$\eta^{RG}$, this does not introduce an additional uncertainty in
our formulae as long as $\eta$ (containing $\eta^{RG}$ as well as
$\eta^{CN}$) is treated as (a small but) unknown parameter. In
this sense, the formula for $\eta^{RG}$ in Eq.~(\ref{etaRG})
should be used only as an estimate for the approximate size of
this correction parameter. The remaining leading log error stems
from the running of the other parameters and is comparatively
small (for hierarchical neutrinos). We also note that since our
formulae only depend on the ratio $\frac{m_2^\nu}{m_3^\nu}$, part
of the leading log error from the running of $m_2^\nu$ and
$m_3^\nu$, due to flavour-blind interactions, cancels out.

For convenience, we summarise the conditions under which the perturbative formulae presented in Eqs.~(\ref{F1}) - (\ref{F3}) can be applied:
\begin{itemize}
\item $\theta_{12}^e$, $\theta_{23}^e$, $\theta_{13}^\nu$ and $\eta$ are small, such that an expansion in these parameters is justified.
\item $\theta_{13}^e$ can be neglected (which is motivated by classes of GUT models where $\theta_{13}^e \approx \lambda^3$).
\item The light neutrino masses are hierarchical, i.e.\ $m_1 \ll m_2 < m_3$.
\item RG and CN corrections are dominated by third family effects (which allows them to be subsumed into the single parameter $\eta= \eta^{RG}+\eta^{CN}$).
\end{itemize}

\section{Deviation parameters}\label{deviation}

Another parametrisation of the lepton mixing matrix can be
achieved by taking an expansion about the TB matrix
\cite{Li:2004dn,King:2007pr,Pakvasa:2007zj,Parke}. Following
\cite{King:2007pr} three small parameters
$r$, $s$ and $a$ may be introduced to describe the deviations of the
reactor (r), solar (s) and atmospheric (a) angles from their TB values:
\begin{equation}\label{de}
s_{13}= \frac{r}{\sqrt{2}}, \hspace{4mm}  s_{12} =
\frac{1}{\sqrt{3}} (1+ s), \hspace{4mm} s_{23} =
\frac{1}{\sqrt{2}} (1+ a)\;.
\end{equation}
Global fits of the conventional lepton mixing angles
\cite{Schwetz:2008er} can be translated into the 2$\sigma$-ranges\footnote{Note that $r$ must be positive definite,
while $s,a$ can take either sign. Indeed there is a preference for $s$ to be negative.}
\begin{equation}\label{bf}
0 < r  <  0.28, \hspace{4mm} -0.10 < s < 0.02, \hspace{4mm} -0.12
< a < 0.12\;.
\end{equation}
The empirical smallness of the parameters $r,s,a$ shows that this
parametrisation is as general as the Wolfenstein parametrisation of the
quark mixing matrix.

Without loss of generality,
the perturbative formulae in Eqs.~(\ref{F1}) - (\ref{F3}) may then be recast in
terms of the deviation parameters as:
\bea \label{F1s}
a&\approx&\left
|1+a^{\nu}+\frac{\eta}{4}-\theta_{23}^{e}e^{i(\delta_{23}^{\nu}-\delta_{23}^{e})}
\right|-1\;,
\\
\label{F2s}
r&\approx&
\left|\theta_{12}^{e}-\frac{1}{3}\frac{m_{2}^{\nu}}{m_{3}^{\nu}}\eta
e^{i(\delta_{12}^{e}-\delta_{12}^{\nu}+2\delta_{23}^{\nu})}
-r^{\nu}e^{i(\delta_{12}^{e}-\delta_{13}^{\nu}+\delta_{23}^{\nu})}\right|\;,
\\
\label{F3s}
s&\approx&\left
|1+s^{\nu}+\frac{\eta}{6}-\theta_{12}^{e}e^{i(\delta_{12}^{\nu}-\delta_{12}^{e})}
\right|-1\;,
\eea
where the $a^{\nu}$, $s^{\nu}$ and $r^{\nu}$ factors parametrise the would-be small deviation from tri-bimaximal setting in the neutrino sector, in full analogy with Eqs.~(\ref{de}).

Concerning the phases, due to the strong dominance of the first terms on the RHS of Eqs.~(\ref{F1}) and (\ref{F3}) the physical factors $\delta_{12}$ and $\delta_{23}$ are essentially identical to the neutrino sector ones, i.e.\ $\delta_{12}\approx \delta_{12}^{\nu}$ and $\delta_{23}\approx \delta_{23}^{\nu}$. Since there is not such a strongly dominant term in (\ref{F2}), the determination of the $\delta_{13}$ phase requires a more careful treatment.

\section{Tri-bimaximal neutrino mixing and sum rules}\label{TBMsum}

So far the above results assume nothing about the nature of the
underlying neutrino mixing angles, apart from the fact that
empirically they must be close to the TB mixing values (within our
GUT motivated framework). Now we shall explore the possibility
that the underlying neutrino mixing angles take TB values quite
accurately, which corresponds to setting
$r^{\nu}=s^{\nu}=a^{\nu}=0$. This situation occurs for example in
models based on certain family symmetries such as $A_4$ or $\Delta_{27}$
\cite{Altarelli:2005yp,Ma:2001dn,Altarelli:2006kg,deMedeirosVarzielas:2005ax,deMedeirosVarzielas:2005qg,King:2006np,deMedeirosVarzielas:2006fc,Altarelli:2008bg,Chen:2007afa}.
Although the neutrino mixing angles may very accurately have the
TB form, the physical lepton mixing angles will still continue to
receive charged lepton and RG/CN corrections, so the observable
lepton mixing angles are not expected to be accurately of the TB
form even in this case.

Setting $r^{\nu}=s^{\nu}=a^{\nu}=0$ and neglecting
$\theta_{13}^{\nu}$ one can use formulae (\ref{F2}) and (\ref{F2s}) to get:
\bea
r&\approx & \sqrt{{\theta_{12}^{e}}^{2}+\left(\frac{1}{3}\frac{m_{2}^{\nu}}{m_{3}^{\nu}}\eta\right)^{2}-\frac{2}{3}\theta_{12}^{e}\frac{m_{2}^{\nu}}{m_{3}^{\nu}}\eta\cos(\delta_{12}^{e}-\delta_{12}^{\nu}+2\delta_{23}^{\nu})}\label{F2f}
\eea
for the reactor angle deviation and\footnote{Here we present formulae for both $\sin\delta$ and $\cos\delta$ to provide a complete information on the Dirac CP phase including the quadrant ambiguity that might arise if only the latter was present.}
\bea
\sin\delta & \approx & \frac{1}{r}\left[{\theta_{12}^{e}\sin(\delta_{12}^{e}-\delta_{12}^{\nu}+\pi)-\frac{1}{3}\frac{m_{2}^{\nu}}{m_{3}^{\nu}}\eta \sin 2\delta_{23}^{\nu}}\right]\label{sin}\;,\\
\cos\delta & \approx &  \frac{1}{r}\left[{\theta_{12}^{e}\cos(\delta_{12}^{e}-\delta_{12}^{\nu}+\pi)+\frac{1}{3}\frac{m_{2}^{\nu}}{m_{3}^{\nu}}\eta\cos 2 \delta_{23}^{\nu}}\right]\label{cos}
\eea
for the lepton sector Dirac CP phase, where $\delta=\delta_{13}-(\delta_{12}+\delta_{23})$  and $\delta_{12}\approx \delta_{12}^{\nu}$, $\delta_{23}\approx \delta_{23}^{\nu}$ and formula (\ref{F2}) have been used. Recall also that with a good accuracy $2\delta_{23}^{\nu}\approx \alpha_{2}$.

In the same case, Eqs.~(\ref{F1s}) and (\ref{F3s}) become:
\bea \label{F1f}
a&\approx& \frac{\eta}{4}-\Delta\;,\\
\label{F3f} s&\approx&  \frac{\eta}{6}+
\theta_{12}^{e}\cos(\delta_{12}^{e}-\delta_{12}^{\nu}+\pi)\;,
\eea
where $\Delta\equiv
\theta_{23}^e\cos(\delta_{23}^{\nu}-\delta_{23}^{e})$ has been used to parameterize the lack of information on the $\delta_{23}^{\nu}-\delta_{23}^{e}$ phase difference.

Notice that formulae (\ref{cos}), (\ref{F1f}) and (\ref{F3f}) constitute a system of three linear equations for three a-priori unknown quantities $\eta$, $\cos(\delta_{12}^{e}-\delta_{12}^{\nu}+\pi)$ and $\Delta$ with coefficients that can be inferred from experiment. If some of these quantities happen to be negligible, the system becomes overconstrained and one can obtain non-trivial relations between the neutrino sector observables.

\subsection{Application to GUT Models}
In this subsection we apply the above results to the case of a specific but well
motivated class of GUT models. More general application of our results will be considered
in the next subsection.

We have previously noted that the typical GUT scale
expectation for the charged lepton mixing
angles is $\theta^e_{12}\approx \lambda/3$,
$\theta^e_{23}\approx \lambda^2$, $\theta^e_{13}\approx
\lambda^3$, where $\lambda \approx 0.22$ is the Wolfenstein
parameter governing the quark mixing matrix.
Numerically this implies that in such a case
$\theta^e_{12}\approx 0.07$,
$\theta^e_{23}\approx 0.05$, $\theta^e_{13}\approx 0.01$.
This provides a justification for neglecting
$\theta^e_{13}$ but keeping both $\theta^e_{12}$ and $\theta^e_{23}$.
The parameter $\eta$ is quite model dependent, in part because of involving the highly model
dependent piece $\eta^{CN}$. Even the contribution from $\eta^{RG}$ can have quite a range
of values, however for hierarchical neutrino masses, assuming the dominant contribution
to arise from the first term in Eq.~(\ref{etaRG}) (i.e.\ $\eta^{RG}=\frac{y_{\tau}^2}{8\pi^2}\ln \frac{M_{GUT}}{M_Z}$),
we may estimate $\eta^{RG}\approx 0.02-0.10$ corresponding to a range of tau Yukawa couplings of
$y_{\tau}= 0.23-0.51$.
In SUSY models this maps to a range of the ratio of Higgs vacuum expectation values (VEVs) $\tan \beta \approx 30-50$, cf.~\cite{Antusch:2008tf}.
Therefore in such typical GUT models, with small CN corrections, we may expect that the quantities
$\eta, \theta_{12}^e$ and $\Delta$ may all be of a similar order of magnitude, with none of them being
negligible. In some sense this is a ``worst case'' situation, since it involves all three quantities,
but on the other hand in this scenario there are theoretical reasons to expect that all of these
quantities are quite small, each giving a correction of less than 10 per cent, which justifies our perturbative approach
(e.g.\ the higher order corrections which we neglect would account for around a per cent or so).

In addition the ratio of assumed hierarchical neutrino masses is given by
$\frac{m_{2}^{\nu}}{m_{3}^{\nu}}\approx \sqrt{\frac{\Delta m_{\odot}^{2}}{|\Delta m_{A}^{2}|}}\approx 0.18$
and the combination which enters the formulae is $\frac{1}{3}\frac{m_{2}^{\nu}}{m_{3}^{\nu}}\approx 0.06$.
In such a GUT motivated framework, with small CN corrections, it is seen that the second and third term
in the square root on the right-hand side of Eq.~(\ref{F2f}) only give a small correction compared to the first term,
since the product $\frac{1}{3}\frac{m_{2}^{\nu}}{m_{3}^{\nu}}\eta$
is much smaller than $\theta^e_{12}$, and in this case we would have the prediction:
\be
r\approx \lambda /3 \approx 0.073
\ee
accurate to about 10 per cent assuming $\theta^e_{12}\approx \lambda/3$ in the considered class of GUT models.
This corresponds to $s_{13}=|U_{e3}|=0.05$ or $\theta_{13}\approx 3^o$ to an accuracy of about 10 per cent.

The above prediction relies on the assumption that $\theta^e_{12}\approx \lambda/3$. This is quite
a strong assumption, since there may be alternative ways of reconciling the down quark masses and
charged lepton masses other than the Georgi-Jarlskog approach. In fact, the GUT scale values of the down quark and charged lepton masses show a strong dependence on possible supersymmetric threshold corrections, as has been pointed out recently in \cite{Antusch:2008tf,Ross:2007az}, which might open up new possibilities \cite{Antusch:2008tf}.
Using Eqs.~(\ref{cos}) and (\ref{F3f}) we may eliminate
$\theta^e_{12}\cos(\delta_{12}^{\nu}-\delta_{12}^{e}+\pi)$
in favour of $\cos\delta$, $\eta$ and $r$ to obtain a sum rule:
\be \label{sum0}
s\approx r\cos\delta + \eta \left(\frac{1}{6}-\frac{1}{3}\frac{m_{2}^{\nu}}{m_{3}^{\nu}}\cos\alpha_{2}\right).
\ee
Unlike the previous case we have chosen to keep the term proportional to the
product $\frac{1}{3}\frac{m_{2}^{\nu}}{m_{3}^{\nu}}\eta$. Even though it is small,
in this case it may conspire with the numerical value $1/6 \approx 0.17$, depending on the phases,
to give a significant effect. In this case we estimate that the
second term on the right-hand side proportional to $\eta$ may give
a correction of up to about 0.02 which is significant compared to the first term which is governed by $r\approx 0.07$.
On the other hand, the second term might involve a partial cancellation of the two terms in the bracket\footnote{A similar effect has been observed in \cite{Boudjemaa:2008jf} where the RG stability of various lepton sector sum-rules is studied.}
and consequently be much smaller than the first term,
in which case we would recover the approximate relation \cite{King:2007pr}
$s \approx r \cos{\delta}$ which is a simple expression of the well known
sum rule \cite{King:2005bj,Masina:2005hf,Antusch:2005kw}:
\be
\label{sumrule}
\theta_{12} - 35.26^o \approx \theta_{13} \cos \delta .
\ee

Finally we remark that it is difficult to predict the atmospheric deviation parameter due to the
unknown phases in the quantity $\Delta\equiv \theta_{23}^e\cos(\delta_{23}^{\nu}-\delta_{23}^{e})$,
plus the RG correction, however in the GUT motivated cases described above we would expect
typically $a\leq 0.1$.

\subsection{Other Limits, Applications and Sum Rules}
In this subsection we consider other limits which are not directly motivated by the GUT-inspired assumptions of the previous subsection.
In the context of more general models of neutrino masses and mixings (satisfying the conditions of section \ref{sec:formulae}) the formulae in Eqs.~(\ref{F1s}) - (\ref{F3s}) can still be applied to analyse under which conditions a precise measurement of the leptonic mixing angles in future neutrino oscillation experiments could provide hints that the underlying neutrino mixing angles indeed satisfy the tri-bimaximal mixing  pattern.

Of course, if all three deviation parameters $\eta, \theta_{12}^e$ and $\Delta$ are negligibly small, tri-bimaximal neutrino mixing would be directly testable. However,
even in the presence of corrections (a situation which is typical in GUT models of flavour) the pattern of leptonic mixing angles may point towards to tri-bimaximal neutrino mixing at some high scale (flavour scale).
The simplest possibility would be that only one of the corrections is important, while the other two can be neglected. This leads to three cases:
\begin{itemize}

\item {\em Only $\eta$ is relevant:} In this case Eqs.~(\ref{F1s}) - (\ref{F3s}) simplify to
\be
a\approx \frac{\eta}{4}\;,\qquad
r\approx\frac{1}{3} |\eta| \frac{m_{2}^{\nu}}{m_{3}^{\nu}} \;,\qquad
s\approx\frac{\eta}{6}\;.
\ee
We note that when non-zero $\theta_{13}$ at low energy is generated only by third family RG (and CN) effects, Eqs.~(\ref{F2f}), (\ref{sin}) and (\ref{cos}) determine the Dirac CP phase $\delta$ to be $\delta = - \alpha_{2}$ for $\eta > 0$ (which is the case for pure RG corrections) or $\delta = - \alpha_{2} + \pi$ for $\eta < 0$ (c.f.~\cite{Antusch:2003kp}).
With three predictions of $a,r$ and $s$ depending only on one parameter $\eta$, there are now two correlations which would provide a ``smoking gun'' signal of an underlying tri-bimaximal neutrino mixing pattern, e.g.\
\be
s \approx \frac{2}{3} a  \; , \qquad
r \approx 2 |s| \frac{m_{2}^{\nu}}{m_{3}^{\nu}} \approx \frac{4}{3} |a| \frac{m_{2}^{\nu}}{m_{3}^{\nu}}\; .
\ee

\item {\em Only $\theta_{12}^e$ is relevant:}
In this limit Eqs.~(\ref{F1s}), (\ref{F2s}) simplify into
\be
a\approx 0 \;,\qquad
r\approx \theta_{12}^{e}\;.
\ee
Moreover, in this limit one gets $\delta_{12}^{\nu}-\delta_{12}^{e}+\pi=\delta$ from (\ref{sin}) and  (\ref{cos}) which also yields:
\bea
s&\approx&
\theta_{12}^{e}\cos{\delta} \;.
\eea
To start with, the first relation $a \approx 0$ (almost exactly maximal mixing $\theta_{23}$) would indicate that (barring cancellations between $\Delta$ and $\eta$ corrections) both $\Delta$ and $\eta$ are negligible \cite{Antusch:2004yx}.
The correlation between the other two corrections can be written as \cite{King:2007pr}
\be \label{sum1}
s \approx r \cos{\delta}\;.
\ee
which is again a compact expression of the sum rule \cite{Antusch:2005kw,King:2005bj,Masina:2005hf}
in Eq.~(\ref{sumrule}).

\item {\em Only $\Delta$ is relevant:}
Although not a typical
situation in flavour models, we mention for completeness that this
case shows that there exists a possible correction to
$\theta_{23}^\nu$, namely $\theta_{23}^e$ which perturbs the
tri-bimaximal pattern only in $a$ while leaving $s=r=0$ as a hint
for underlying tri-bimaximality.
\end{itemize}

Let us now turn to the somewhat less simple situation that two of
the corrections are important. With two unknowns and three
measurements, one correlation between the observables remains to
provide a possible signal of tri-bimaximal neutrino mixing. The
three possible cases are as follows:

\begin{itemize}

\item {\em Only $\eta$ and $\theta_{12}^e$ are relevant, $\Delta$ is negligible:}
In the limit that $\Delta = 0$, Eq.~(\ref{F1f}) yields $a\approx \frac{\eta}{4}$ which allows to express $\eta$ in terms of $a$ in the other two equations. Combining them  we find the improved sum rule
\be \label{sum3}
s\approx r\cos\delta + \frac{2}{3}a \left(1-2\frac{m_{2}^{\nu}}{m_{3}^{\nu}}\cos\alpha_{2}\right)
\ee
which, compared to  \cite{Antusch:2007ib,Antusch:2007vw}, includes next-to-leading correction to $s$ of the form  ${\cal O}( a \, \frac{m_2}{m_3})$.
This new term, however, depends on the Majorana phase $\alpha_2$, which will be difficult to measure.

\item {\em Only $\theta_{12}^e$ and $\Delta$ are relevant, $\eta$ is negligible:}
In this case, to leading order in small parameters, we again obtain the sum rule of Eq.~(\ref{sum1}), however now with $a \not= 0$.\footnote{We remark that for this scenario it is also possible to derive a sum rule which is exact in  $\theta_{23}$ \cite{Antusch:2007rk}.}

\item {\em Only $\Delta$ and $\eta$ are relevant, $\theta_{12}^e$ is negligible:}  Again, this is not a typical situation in flavour models (since usually a correction $\Delta$, containing $\theta_{23}^e$, is accompanied by a correction $\theta_{12}^e$), however for completeness we mention that here the correlation
\begin{eqnarray}
r = 2 |s| \frac{m_{2}^{\nu}}{m_{3}^{\nu}}
\end{eqnarray}
remains as a hint for underlying tri-bimaximal neutrino mixing.

\end{itemize}

Finally, if $\eta,\theta_{12}^e$ and $\Delta$ are all important, we have as many observables  as unknowns which means that predictivity is lost.
Combining Eqs.~(\ref{cos}) - (\ref{F3f}) to eliminate $\eta$ and $\theta_{12}^e\cos(\delta_{12}^{\nu}-\delta_{12}^{e}+\pi)$ one arrives to a $\Delta$-dependent sum rule:
\be \label{sum4}
s\approx r\cos\delta + \frac{2}{3}(a+\Delta) \left(1-2\frac{m_{2}^{\nu}}{m_{3}^{\nu}}\cos\alpha_{2}\right).
\ee
This result may be in principle (depending again on the hard-to-measure Majorana CP phase $\alpha_2$) used to determine $\Delta$, which may then be compared
to the theoretical expectation for $\Delta$ within specific GUT models.

Even in such a case, Eqs.~(\ref{F1s}) - (\ref{F3s}) may be very useful. For example, they can be used
to determine the possible values of the correction parameters $\eta,\theta_{12}^e$ and $\Delta$ under the assumption of underlying nearly exact tri-bimaximal neutrino mixing. This provides a useful information for model building. Eqs.~(\ref{F1s}) - (\ref{F3s}) can also be applied in context of various models of flavour that might happen to predict some of the correction parameters.\footnote{Also, if models predict the Majorana CP phase $\alpha_2$, this would improve the predictivity and testability in some cases.}
Then, predictivity would be regained and one could derive correlations.

\section{Conclusion}\label{concl}

Many GUTs predict two large neutrino mixing angles,
with the underlying charged lepton mixing angles being small and quark-like,
and the neutrino masses being hierarchical. In such frameworks we
present simple approximate analytic formulae giving the lepton
mixing angles in terms of the underlying high energy neutrino
mixing angles together with small perturbations due to both
charged lepton corrections and RG/CN effects.
The resulting analytic formulae given in
in Eqs.~(\ref{F1}) - (\ref{F3}) (or equivalently Eqs.~(\ref{F1s}) - (\ref{F3s}))
express the lepton mixing angles
in terms of the neutrino mixing angles and
the leading order corrections due to $\theta_{ij}^e$ and $\eta$,
which represent the leading order
terms in an expansion in powers of small parameters
representing both charged lepton mixing
corrections and third family RG/CN effects.

We have applied these perturbative formulae to the ubiquitous case of
tri-bimaximal neutrino mixing at the unification scale, in order
to predict the theoretical corrections to mixing angle predictions
and sum rule relations, and have given a general discussion of all
limiting cases. When applied to GUT models, we have seen that
the formulae lead to a roughly 10 per cent correction to reactor angle
prediction based on the Georgi-Jarlskog ansatz of $\theta_{13}\approx 3^o$.
More generally, independently of the GJ ansatz, we have seen that
the sum rule relation $s\approx r \cos \delta$ receives a correction
given by Eq.~(\ref{sum0}) which we estimate to be up to about 0.02
and thus may be significant compared to $r\approx 0.07$.
We have also relaxed the GUT motivated assumptions, and
obtained a variety of other possible relations amongst observable
parameters, which, if confirmed by experiment, could signal
{\it neutrino} tri-bimaximal mixing in a more general context than the GUT paradigm.
For example, if for some reason $\theta_{23}^e$ turned out to be negligible,
then our perturbative formulae lead to the novel relation in Eq.~(\ref{sum3})
which, although being quite accurate, involves the Majorana phase
$\alpha_2$, making it difficult to test.

In conclusion, the perturbative formulae presented here
provide a useful physical insight into the origin and
nature of the deviations of the observable lepton mixing angles from
the underlying {\it neutrino} mixing angles at the GUT scale,
thereby opening a window into the nature of the high energy GUT theory.
The results may also be used to test in a more general way the
hypothesis of tri-bimaximal mixing in the {\it neutrino} sector
under various assumptions about the nature of the
charged lepton and third family RG/CN corrections,
each of which leads to a different testable relation.
In this way the underlying nature of tri-bimaximal {\it neutrino} mixing
(if present) may be revealed in the low energy neutrino experiments
which measure only the physical lepton mixing angles and phases.
In addition, such perturbative formulae
may also be useful in a more prosaic way by
speeding up multi-parameter scans in particular
GUT models, or simply as a means of checking the numerical
results.

\section*{Acknowledgments}
S.A.\ acknowledges partial support from
the DFG cluster of excellence ``Origin and Structure of the
Universe'' and from the grant EU ILIAS RII3-CT-2004-506222.
SFK acknowledges: PPARC Rolling Grant
PPA/G/S/2003/00096; EU Network MRTN-CT-2004-503369; EU ILIAS
RII3-CT-2004-506222. The work of M.M. has been supported from the PPARC Rolling Grant PPA/G/S/2003/00096 and from the Royal Institute of Technology in Stockholm under the Contract Number KTH SII-56510.

\providecommand{\href}[2]{#2}\begingroup\raggedright\endgroup

\end{document}